\documentclass[aps, prd, 10pt, twocolumn, superscriptaddress,noshowpacs, preprintnumbers, longbibliography,nofootinbib,bibnotes,floatfix]{revtex4-2}
\pdfoutput=1 

\usepackage{amsmath}
\usepackage{amsfonts}
\usepackage{amssymb}
\usepackage{mathtools}
\usepackage{graphics}
\usepackage{tabularx}
\usepackage[export]{adjustbox}
\usepackage{multirow}
\usepackage{citesort}
\usepackage{graphicx}
\usepackage{url}
\usepackage{soul}
\usepackage{bm}
\usepackage[dvipsnames]{xcolor}
\usepackage[utf8]{inputenc}
\usepackage[normalem]{ulem}
\usepackage[colorlinks=true,breaklinks=true]{hyperref}
\hypersetup{allcolors=[rgb]{0.0 0.0 1.0},linkcolor=[rgb]{0.75 0.05 0.05}}

\usepackage{tikz,xcolor,hyperref}

\definecolor{lime}{HTML}{A6CE39}
\DeclareRobustCommand{\orcidicon}{\hspace{-1mm}
	\begin{tikzpicture}
	\draw[lime, fill=lime] (0,0) 
	circle [radius=0.16] 
	node[white] {{\fontfamily{qag}\selectfont \tiny \,ID}};
	\draw[white, fill=white] (-0.0525,0.095) 
	circle [radius=0.007];
	\end{tikzpicture}
	\hspace{-3mm}
}

\foreach \x in {A, ..., Z}{\expandafter\xdef\csname orcid\x\endcsname{\noexpand\href{https://orcid.org/\csname orcidauthor\x\endcsname}
			{\noexpand\orcidicon}}
}

\begin{document}

\title{Neutrinos from stars in the Milky Way}
\author{Pablo Mart\'inez-Mirav\'e\orcidA{}}
\email{pablo.mirave@nbi.ku.dk}
\affiliation{Niels Bohr International Academy and DARK, Niels Bohr Institute, University of Copenhagen,\\
Blegdamsvej 17, 2100, Copenhagen, Denmark}

\author{Irene Tamborra\orcidB{}}
\email{tamborra@nbi.ku.dk}
\affiliation{Niels Bohr International Academy and DARK, Niels Bohr Institute, University of Copenhagen,\\
Blegdamsvej 17, 2100, Copenhagen, Denmark}

\begin{abstract}
Neutrinos are  produced during stellar evolution by means of thermal and thermonuclear processes.
We model the cumulative  neutrino flux  expected at Earth from all stars in the Milky Way:  the Galactic stellar neutrino  flux (GS$\nu$F).  We account for  the star formation history of our Galaxy and reconstruct the spatial distribution of Galactic stars  by means of a random sampling procedure based on  Gaia Data Release 2. We use  the stellar evolution code \texttt{MESA} to compute the neutrino emission for a suite of  stellar models with solar metallicity and zero-age-main-sequence mass between $0.08M_\odot$ and $100\ M_\odot$, from their pre-main sequence phase to their final fates. 
We then reconstruct the evolution of the neutrino spectral energy distribution for each stellar model in our suite.
The GS$\nu$F   lies  between  $\mathcal{O}(1)$~keV and $\mathcal{O}(10)$~MeV, with thermal (thermonuclear) processes responsible for shaping neutrino emission at energies smaller (larger) than $0.1$~MeV. Stars with mass larger than $\mathcal{O}(1\ M_\odot)$, located in the thin disk of the Galaxy, provide the largest contribution to the GS$\nu$F. Moreover, most of the GS$\nu$F originates from  stars distant from Earth about $5$--$10$~kpc, implying that  a large fraction of stellar neutrinos can reach us from the Galactic Center. Solar neutrinos and the diffuse supernova neutrino background have energies comparable to those of the GS$\nu$F, challenging the detection of the latter.  However, directional information of solar neutrino and GS$\nu$F  events, together with the annual modulation of the solar neutrino flux, could  facilitate the  GS$\nu$F detection; this will kick off a new era for low-energy neutrino astronomy, also providing a novel probe to discover New Physics. 
\end{abstract}

\maketitle

\section{Introduction}
Stars radiate energy by emitting neutrinos, in addition to electromagnetic emission. 
By means of nuclear fusion in the  core of stars, light atomic nuclei combine to form heavier ones. {\it Thermonuclear neutrinos} are produced as a by-product 
with  $\mathcal{O}(0.1$--$10$)~MeV  energy (see, e.g., Refs.~\cite{Schwarzschild:1958,Dina:2009,Kippenhahn:2012qhp} for an overview of thermonuclear processes).  Stars also emit {\it thermal neutrinos} from  thermal processes that strongly depend on the temperature, density, and composition of the medium;  thermal neutrinos have typical energies of  $\mathcal{O}(10$--$100)$~keV (see e.g.~Refs.~\cite{Itoh:1996,Raffelt:1996wa,Kippenhahn:2012qhp} for an introduction to thermal processes in stars). Hence, all stars in our Universe are neutrino sources, including those in our own Galaxy--the Milky Way. The  flux of Galactic neutrinos from stars  was first estimated in Ref.~\cite{Brocato:1997tu},  relying on the FRANEC evolutionary code~\cite{Castellani:1992pn} and a Galactic model from Ref.~\cite{Bahcall:1986a}. Similar results for thermonuclear neutrinos were presented in Ref.~\cite{Porciani:2003zq} for other models accounting for the   structure and evolution of our Galaxy, relying on the stellar evolution results and neutrino emission properties presented  in Ref.~\cite{Brocato:1997tu}. 

Our understanding of stellar evolution and neutrino production in stellar environments has greatly improved in the past decade; we refer the reader to, e.g., Refs.~\cite{Vitagliano:2019yzm,Tamborra:2024fcd} for recent reviews on the topic.
Solar neutrinos from different thermonuclear channels, such as  the pp-chain~\cite{Super-Kamiokande:2001ljr,SNO:2001kpb,Borexino:2017rsf,Borexino:2017uhp} and  the CNO cycle~\cite{Borexino:2023}, have been experimentally identified. The thermal neutrino emission from the Sun has also been accurately computed~\cite{Vitagliano:2017odj}. In this regard, solar neutrino spectroscopy has been fundamental to improve our understanding of the Sun, see e.g.~Refs.~\cite{Bahcall:1989ks,Bahcall:1992hn, Bahcall:2004pz,Haxton:2012wfz,Gann:2021ndb}, as well as to learn about neutrino properties~\cite{Maltoni:2015kca,Wurm:2017cmm}. In addition, we have gained a deeper understanding of the  dependence of neutrino emission on  stellar mass and metallicity~\cite{Farag:2020nll,Farag:2023xid,Patton:2017neq}.

\begin{figure*}
    \centering
    \includegraphics[width=\linewidth]{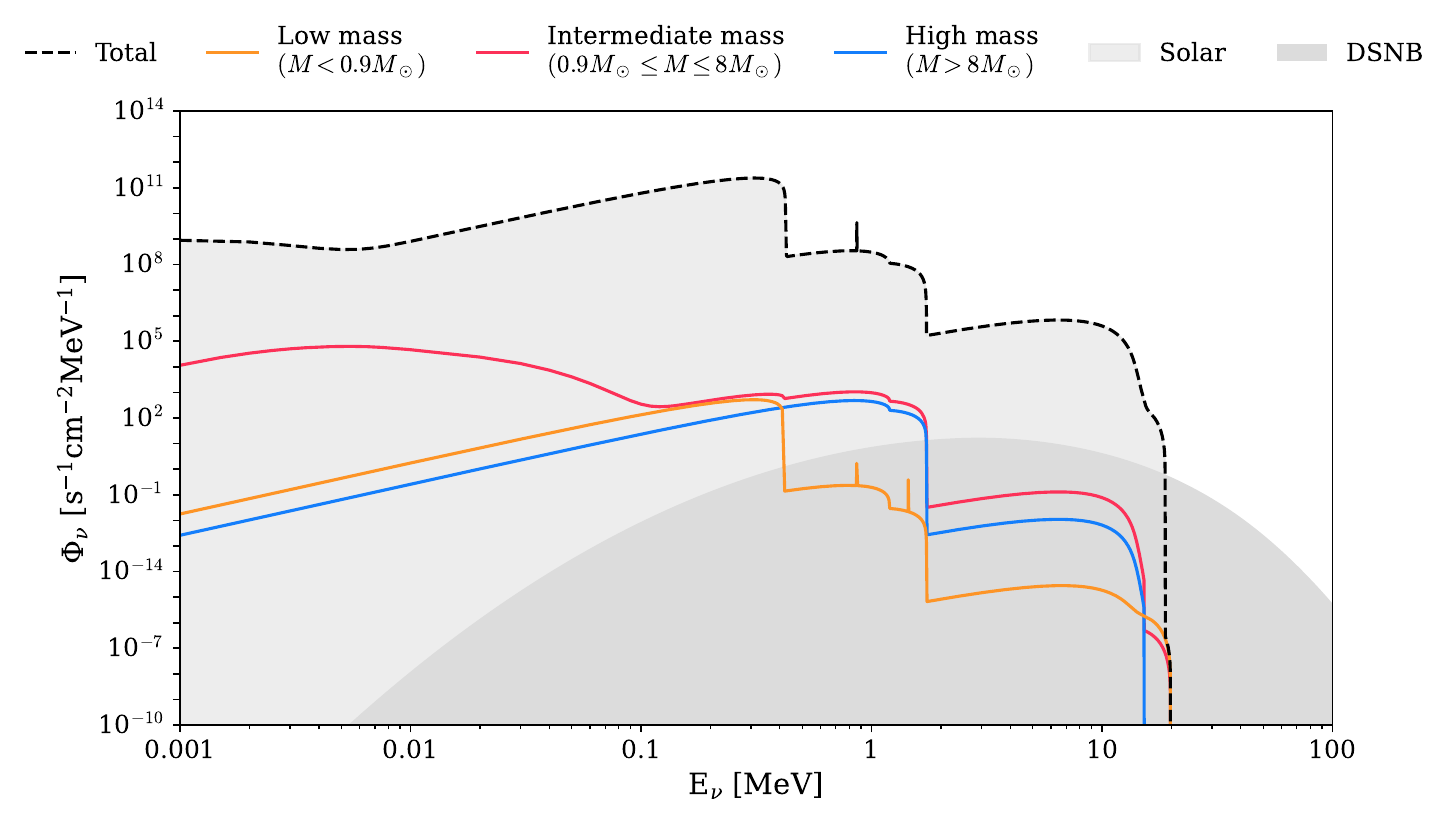}
    \caption{Galactic stellar neutrino and antineutrino flux  at Earth for all flavors as a function of neutrino energy (dashed line), including solar neutrinos. The contributions from low-mass  ($0.08 M_\odot  \lesssim M < 0.9 M_\odot)$, intermediate-mass  ($0.9 M_\odot \leq M \lesssim 8 M_\odot$), and high-mass  ($M \geq 8 M_\odot$) stars are  displayed as solid lines in orange, red, and blue, respectively. The light and dark gray shaded areas indicate the flux expected at Earth from solar neutrinos and the  DSNB, respectively, as from Ref.~\cite{Vitagliano:2019yzm}.}
    \label{fig:spectrum}
\end{figure*}

In this work, we forecast of the Galactic stellar neutrino flux (GS$\nu$F): the cumulative neutrino flux from all  stars in the Milky Way expected at Earth. Our findings are summarized in Fig.~\ref{fig:spectrum}. We rely on Ref.~\cite{Cautun:2020}, that employs the Gaia Data Release 2, for modeling  the spatial distribution of stars in our Galaxy and use new insights  from galactic archaeology~\cite{Matteucci:2021a} to model the Galactic star formation history (SFH). Moreover, we compute the neutrino emission from stars with zero-age-main-sequence (ZAMS) mass between $0.08M_\odot$ and $100\ M_\odot$, following the stellar evolution from the pre-main sequence stage to the final evolutionary stages. 

This paper is organized as follows. In Sec.~\ref{sec:mw}, we describe our modelling of the Milky Way stellar population, including the spatial distribution of stars and the SFH. In Sec.~\ref{sec:evolution}, we outline  our modelling of the evolution of each star in our suite. Section~\ref{sec:emission} describes the approach adopted to compute the GS$\nu$F.  Section~\ref{sec:flux} presents our results for  low-mass stars ($M < 0.9 M_\odot$), intermediate-mass stars ($0.9 M_\odot \lesssim M < 8 M_\odot$), and high-mass stars ($M \geq 8M_\odot$); the main  GS$\nu$F features are also discussed. Finally, we summarize our findings and discuss the GS$\nu$F detectability prospects in Sec.~\ref{sec:conclusion}.

\section{Galactic star formation rate}
\label{sec:mw}
In this section, we present the model of the Galactic stellar population  and our parametric description of the birthrate of stars.

\begin{figure*}
    \centering
    \raisebox{-0.5\height}{\includegraphics[width=0.495\linewidth]{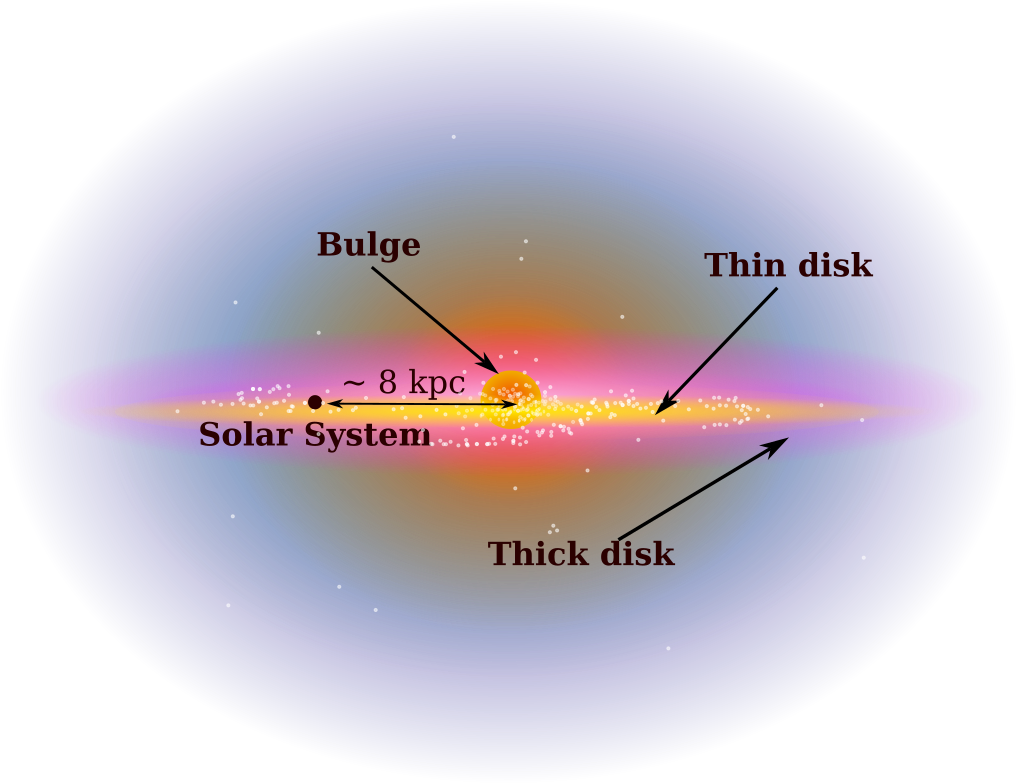}}
    \raisebox{-0.5\height}{\includegraphics[width=0.495\linewidth]{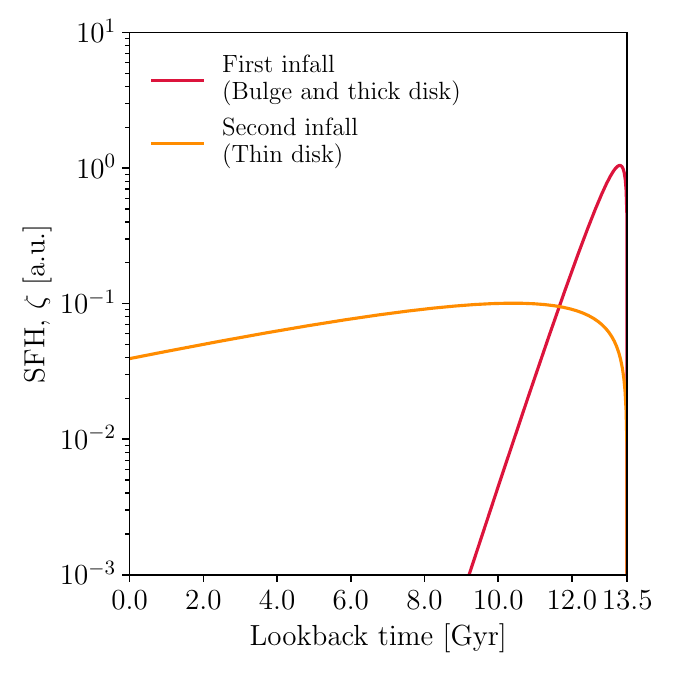}}
    \includegraphics[width = \linewidth]{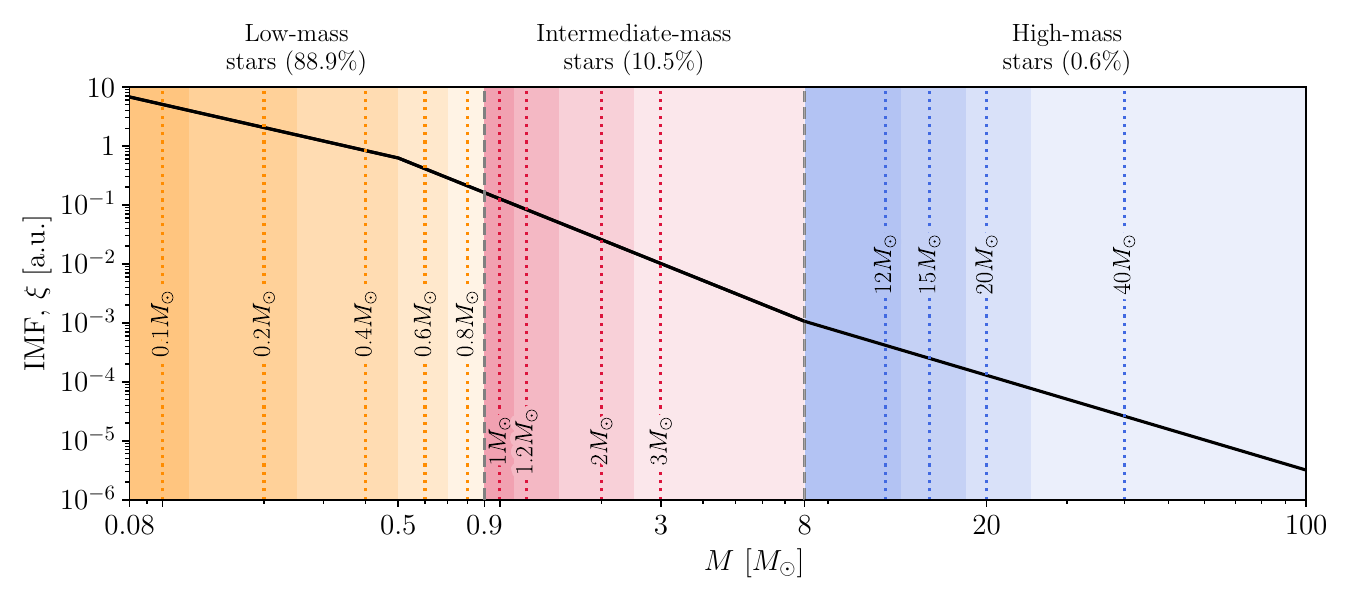}
    \caption{{\it Top left:} Sketch of the Milky Way, highlighting its three components: the bulge, the thin disk, and thick disk. To guide the eye, we also mark  the distance of the Solar System from the Galactic Center. {\it Top right:} SFH  of the Milky Way ($\zeta$, Eq.~\ref{eq:sfh}), normalized to unity and in  in arbitrary units (a.u.), as a function of the lookback time, assuming the Universe is $13.5$~Gyr old. We assume the two-infall  formation model. The red line corresponds to the first infall episode, responsible for the formation of the bulge and thick disk; the orange line represents the second infall episode, leading to the formation of the thin disk. {\it Bottom:} IMF  ($\xi$, in black, Eq.~\ref{eq:imf})  normalized to unity,   in arbitrary units, and as a function of the ZAMS mass. The vertical dashed lines indicate the ZAMS mass of the progenitors evolved in {\tt MESA} in this work; the shaded bands indicate the range of ZAMS masses for which each model, marked by a dashed line, is assumed to  be representative. The orange, red, and blue bands correspond to our  low-  ($88.9\%$), intermediate-  ($10.5\%$), and high-mass  ($0.6\%$)  stellar populations.}
    \label{fig:model}
\end{figure*}

\subsection{Spatial distribution of stars}
As sketched in the top left panel of Fig.~\ref{fig:model}, we divide the baryonic mass distribution in our Galaxy  in three independent components: the Galactic buldge, a thin disk, and a thick disk. We describe the mass distribution of stars in the Milky Way in  Galactocentric coordinates $(R, z)$. 

For the bulge, we adopt a parameterization of the baryonic density that does not account for the complex stellar distribution in the bulge and neglects the bar structure~\cite{Bissantz:2002,Mcmillan:2017}:
\begin{align}
    \rho_{\rm bulge} = \frac{\rho_{0, \textrm{bulge}}}{(1 + r'/r_0)^\alpha} \exp \left[-(r'/r_{\rm cut})^2\right]\, , 
    \label{eq:bulge}
\end{align}
with $r' = \sqrt{R^2 + (z/q)^2}$~\footnote{This parametrization is the axisymmetric version of the model proposed in  Ref.~\cite{Bissantz:2002}.}. Here, $q$, $r_0$, $r_{\rm cut}$, and $\alpha$ are phenomenological parameters determining the three-dimensional shape and size of the bulge, and $\rho_{0,{\rm bulge}}$ is the baryon density of the bulge.

The baryon density of the thin and thick  disks is modeled in three dimensions as follows~\cite{Juric:2008,Pouliasis:2017}:
\begin{align}
    \rho_{d} = \frac{\Sigma_0}{2z_d} \exp\left( -\frac{|z|}{z_d} - \frac{R}{R_d}\right) \, ;
    \label{eq:disk}
\end{align}
 the parameters $z_d$ and $R_d$ determine the size of each of the disks, and $\Sigma_0$ is the corresponding density.

Table~\ref{tab:MWdensity}  summarizes the values adopted for the parameters introduced in Eqs.~\ref{eq:bulge} and \ref{eq:disk}, following  Ref.~\cite{Cautun:2020}. The  parameterization proposed  in Ref.~\cite{Cautun:2020} relies on Gaia Data Release 2~\cite{Gaia}, assumes a Navarro-Frenk-White dark matter distribution~\cite{Navarro:1995iw},  an axisymmetric model of the Milky Way with no spiral arms, and accounts for two additional gas disks as well as the circumgalactic medium. Under these assumptions, the baryonic mass of the bulge is $M_{\rm bulge} = 9.2 \times 10^9 M_\odot$, the baryonic mass of the thin disk is $M_{\rm d, thin} = 3.98 \times 10^{10} M_\odot$, and the one of the thick disk is $M_{\rm d, thick} = 1.07 \times 10^{10} M_\odot$~\cite{Cautun:2020}. 

We implement the arm structure of the Milky Way of the thick and thin disks, specifically the arms Scutum–Centaurus, Sagittarius–Carina, Perseus, and Norma-Outer following Ref.~\cite{Denyshchenko:2024}. However, we do not account for the gas disks and the intergalactic medium. Those additional components have a small  density of stars and, therefore, are expected to provide  a negligible contribution to the GS$\nu$F. The uncertainties in the morphology and baryonic mass of the Galactic components (see e.g.~Ref.~\cite{Iocco:2015xga}) can further affect our forecast.
\begin{table}[]
    \centering
    \renewcommand{\arraystretch}{1.6}
    \caption{Parameters used to describe the  contributions to the baryonic mass of the Milky Way, modelled following Eqs.~\ref{eq:bulge} and \ref{eq:disk}, as from Ref.~\cite{Cautun:2020}.} 
    \begin{tabular}{ccc}
         \hline\hline Contribution &Quantity & Value  \\ \hline\hline
         \multirow{5}{*}{Bulge} & $r_0$ & 75 pc  \\
         & $r_{\rm cut}$ & 2.1 kpc  \\
         & $\alpha$ & 1.8 \\
         & $q$ & 0.5  \\
         &$\rho_{0, \textrm{bulge}}$& $101\, M_\odot\, {\rm pc}^{-3}$ \\ \hline
         \multirow{3}{*}{Thin disk} & $z_{d, \textrm{thin}}$ & 300 pc  \\
         &$R_{d, \textrm{thin}}$ & $2.43$ kpc  \\
         &$\Sigma_{0,\textrm{thin}}$ & $1070\,\,M_\odot\, {\rm pc}^{-2}$  \\ \hline
         \multirow{3}{*}{Thick disk} & $z_{d, \textrm{thick}}$ & 900 pc  \\
         &$R_{d, \textrm{thick}}$ & $3.88$ kpc \\
         &$\Sigma_{0, \text{thick}}$ & $113\,M_\odot\, {\rm pc}^{-2}$ \\ \hline\hline
    \end{tabular}
    \label{tab:MWdensity}
\end{table}

\subsection{Birthrate function}
The stellar birthrate function, $B$, provides the number of stars born in the time interval $(t, t+{\rm d}t)$ and  mass interval $(M, M+{\rm d}M)$ in the Milky Way. It can be expressed as the product of the SFH ($\zeta(t)$) and the initial mass function (IMF, $\xi(M)$):
\begin{align}
    B(t,M)\,{\rm d}t \,{\rm d}M= \zeta(t)\,\xi(M)\,{\rm d}t \,{\rm d}M\, ;
\end{align}
note that  we  assume that the IMF is constant in time and the star formation rate does not depend on $M$. 

\subsubsection{Star formation history}
\label{sec:sfh} 
Different methods have been proposed to  investigate the star formation rate of   the Galaxy based on its  chemical evolution, such as the {\it serial} and  {\it parallel approaches}~\cite{Matteucci:2021a}; in the former, the formation of the Milky Way is modeled through  continuous accretion of
gas during which the halo, thick and thin disks are formed, or assuming several episodes of gas infall. The latter approach considers that Galaxy formation results from different episodes of gas accretion occurring in parallel, but at a different rate.

In this work, we adopt an effective two-infall model that parametrizes the star formation rate of the Milky Way following Ref.~\cite{Chiappini:1996ws}~\footnote{Alternative effective parametrizations of the SFH  have been proposed in the literature, including a decreasing exponential function~\cite{Simha:2014rta} or a Gaussian distribution~\cite{Snaith:2015}.}. 
The two-infall model assumes that an initial collapse led to the formation of  the halo  of the thick disk. Star formation continued at a very efficient rate (much larger than the current one) until the gas density dropped below a certain threshold. The gas lost by the halo accumulated in the center, forming the bulge (cf.~left top panel of Fig.~\ref{fig:model}). 
Then, a second infall event is responsible for the thin disk. This  was either the result of a merger event with a small galaxy or  due to the longer time required for material with a high angular momentum to accrete. A  recent analysis supports the occurrence of a delayed episode of gas infall~\cite{Spitoni:2019a}.

We consider the first infall epoch, which resulted in the formation of the bulge and the thick disk, and a later infall forming the thin disk (see the  top left panel of Fig.~\ref{fig:model}). We parametrize the SFH  during the two-infall episodes as
\begin{align}
    \zeta(t) \propto t^\kappa \exp(- t/\tau)\, ,
    \label{eq:sfh}
\end{align}
and determine the effective parameters $\kappa$ and $\tau$ for each infall episode by fitting  the two-infall best-fit model from Ref.~\cite{Grisoni:2017a}.  For the SFH of the bulge and thick disk, we take $\kappa =0.4$ and $\tau = 0.5$~Gyr, while for the  thin disk SFH, we assume $\kappa=0.5$ and $\tau = 6.2$~Gyr. 

The top right panel of Fig.~\ref{fig:model} shows the SFH for each of the two infalls, using the parameters above. Stars in the bulge and thick disk formed rapidly ($12$--$13$~Gyr ago), followed by  stars in the thin disk forming at a roughly constant rate for the past $12$~Gyr. Hence, the bulge and thick disk host older stars, whereas the stellar population of the thin disk  should exhibit a wider spread in stellar age.

\subsubsection{Initial mass function}
The IMF  describes the distribution of stars at birth as a function of stellar mass (see, e.g., Ref.~\cite{Kroupa:2024nax} for a recent review). We adopt a broken power-law for the IMF~\cite{Kroupa:2000iv}:
\begin{align}
\label{eq:imf}
    \xi(M) &\propto M^{-\alpha} \\&\text{with}\begin{cases}
        \alpha = 1.3 \qquad 0.08\ M_\odot\leq M  < 0.5\ M_\odot \\
        \alpha = 2.3 \qquad 0.5\ M_\odot \,\,\,\leq M \leq 100\ M_\odot \\
    \end{cases}\, .
\end{align}
The lower limit in mass of the IMF is chosen according to the fact that, if a proto-star forms with a mass smaller than $0.08\ M_\odot$, its internal temperature never becomes high enough for thermonuclear fusion to begin, leading to a brown dwarf. The choice of the IMF upper limit in this work is arbitrary, since  the IMF strongly suppresses the population of high-mass stars. The bottom panel of Fig.~\ref{fig:model} illustrates   the IMF in Eq.~\ref{eq:imf}. 

\section{Stellar evolution}
\label{sec:evolution}
In this section, we present our modeling of the evolution of stars with ZAMS mass of $[0.08, 100]\ M_\odot$. We also illustrate the approach adopted to reconstruct the spectral energy distribution of neutrinos.

\subsection{Stellar models}
\label{subsec:solar-calibration}

\subsubsection{Stellar population modeling} 
We construct a suite of stellar models, considering  each star with fixed ZAMS mass between $[0.1, 100]\ M_\odot$ as a single, non-rotating object with solar metallicity. 
We divide the population of stars in three groups:  {\it low-mass stars} with $M < 0.9 M_\odot$, \textit{intermediate-mass stars} with $0.9\,M_\odot\leq M < 8\,M_\odot $, and {\it high-mass stars} with $ M \geq 8M_\odot$. 

As illustrated in the bottom panel of Fig.~\ref{fig:model},  we select a discrete subset of $13$ progenitors with ZAMS  $M/M_\odot = 0.1$, $0.2$, $0.4$, $0.6$, $0.8$, $1$, $1.2$, $2$, $3$, $12$, $15$, $20$, and $40$. We evolve each of these models using \texttt{Modules for Experiments in Stellar Astrophysics} (\texttt{MESA}) 24.08.1~\cite{Paxton:2011,Paxton:2013pj,Paxton:2015jva,Paxton:2019,MESA:2022zpy}, adapting existing models from the \texttt{MESA} Test Suite.
We assume that each of these progenitors is representative of a range of ZAMS masses, as shown in the bottom panel of Fig.~\ref{fig:model}. The models with mass $\lesssim 1\ M_\odot$ are evolved from their pre-main sequence evolutionary stage until their terminal age main sequence (i.e., until hydrogen is approximately exhausted, since their main-sequence lifetime  would otherwise be  larger than the age of the Universe). Intermediate-mass progenitors ($1\ M_\odot \leq M \leq 3\ M_\odot$) are evolved until the white-dwarf phase, when the photon luminosity significantly exceeds  the neutrino luminosity. Finally, the high-mass progenitors are evolved until pre-core-collapse conditions are reached.

\subsubsection{Convection} 
For each stellar model, we adopt a mixing-length treatment of convection, following Refs.~\cite{Henyey:1965, Paxton:2011}. We fix the mixing-length parameter ($\alpha_{\rm ML}$), relying on a solar calibration. Other parameters related to the mixing-length velocity and temperature gradient in a rising bubble are set to default as in Ref.~\cite{Paxton:2011}. We also account for convective overshooting with an exponential parametrization of the diffusion coefficient, as illustrated in Ref.~\cite{Herwig:2000sq}, with $f_{\rm{ov}} = 0.016$~\cite{Farag:2020nll}~\footnote{Note that the treatment of convection might not be ideal for the modelling of stars with ZAMS mass far from $1\ M_\odot$. We comment on the impact of this caveat in Sec.~\ref{sec:flux}.}.

\subsubsection{Solar calibration} 
We  calibrate our suite of models to solar data  to fix the initial hydrogen, helium, and heavy-element mass fraction (X, Y, and Z, respectively), as well as the mixing length parameter $\alpha_{\rm ML}$. To this end, we use the \texttt{MESA} \texttt{simplex}   built-in routine. We find the set of parameters that best reproduces the solar luminosity ($L_\odot = 3.828 \times 10^{33}$~erg~s$^{-1}$~\cite{IAU:2015fjh}), effective temperature ($T_{\rm eff} = 5772$~K~\cite{IAU:2015fjh}), helium surface fraction ($Y_{\rm surface} = 0.2485 \pm 0.0035$~\cite{Basu:2004}), the radius of the convective zone ($R_{\rm cz} = (0.713 \pm 0.001) R_\odot$~\cite{Basu:1997}), and the measured sound-speed profile~\cite{Rhodes:1997}.

We perform calibration for two  solar models with different predicted surface abundances of heavy elements: $(Z/X)_{\rm surface} = (1.8 \pm 0.1)\times 10^{-2}$ for the AGSS09 solar model~\cite{Asplund:2009fu} and $(Z/X)_{\rm surface} = (2.3 \pm 0.1)\times 10^{-2}$ for the GS98 solar model~\cite{Grevesse:1998bj}, for the estimated age of the Sun ($4.61 \times 10^9$~Gyr)~\cite{Bahcall:1995bt,Bahcall:2005va}.  We have tested that the predicted neutrino flux from the pp-chain, which dominates the solar neutrino emission, is in agreement with observations for both  solar models. Hereafter, we adopt  the calibration values for the GS98 model ($X_0 = 0.7020$, $Y_0 = 0.2778$, $Z_0 = 0.202$, and $\alpha_{\rm ML} = 1.96$), which yield a solar neutrino luminosity $L_{\nu, \odot} = 0.02469 L_{\odot}$, where $L_{\odot}$ is the photon luminosity.

\subsubsection{Nuclear network, thermonuclear and thermal processes} 
We rely on the built-in \texttt{MESA} nuclear reaction networks, \texttt{mesa\_49} for low- and intermediate-mass stars and \texttt{mesa\_206} for the high-mass stars. The nuclear reaction rates are taken from the Joint Institute for Nuclear Astrophysics REACLIB library~\cite{Cyburt:2010} and account for  plasma screening effects~\cite{Chugunov:2007ae} (we refer the interested reader to Refs.~\cite{Paxton:2019, Bauer:2023} for details on the calculation of the nuclear reaction rates). In addition, tabulated weak reaction rates are  from Refs.~\cite{Langanke:2000, Oda:1994, Fuller:1985}. 

Table \ref{tab:reactions} summarizes some of the thermonuclear and thermal reactions, taking place during stellar evolution and that lead to neutrino production. We take these reaction channels into account for computing the GS$\nu$F, as illustrated below. 

\begin{table*}[]
    \centering
    \caption{Thermonuclear and thermal neutrino-emitting reactions considered in the computation of the GS$\nu$F. The last column indicates the stage of stellar evolution during which each reaction channel dominates (if applicable). Notice that all thermonuclear reactions emit solely electron neutrinos. For thermal neutrinos, neutrino-antineutrino pairs are  produced. The processes marked by an asterisk are not included in the computation of the GS$\nu$F; see main text for details.}
    \renewcommand{\arraystretch}{1.4}
    \begin{ruledtabular}
    \begin{tabular}{lccc}
    \multicolumn{4}{c}{Thermonuclear reactions}  \\ \colrule
    \multirow{5}*{pp chain} & pp & $p+p\to ^2{\rm H} + e^+ +\nu_e$ & \multirow{5}*{\shortstack{Dominant for  $T \lesssim 1.9 \times 10^7 $~K,\\ in main-sequence low-mass stars.}} \\
     & pep  & $p + e^- + p\to ^2{\rm H} +\nu_e$& \\
     & $^7$Be   & $^7{\rm Be} + e^- \to ^7{\rm Li}+\gamma+\nu_e$ & \\
     & $^8$B  & $^8{\rm B} \to 2\alpha + e^+ + \nu_e$ & \\
     & hep  & $^3{\rm He}+p\to^4{\rm He}+e^++\nu_e$& \\
     \\
    \multirow{3}*{CNO cycle} & $^{13}$N  & $^{13}{\rm N}\to^{13}{\rm C} + e^+ + \nu_e$ & \multirow{3}*{\shortstack{Dominant for  $T \gtrsim 1.9 \times 10^7$~K,\\ in main sequence intermediate- \\ and high-mass stars.}} \\
     & $^{15}$O  & $^{15}{\rm O} \to ^{15}{\rm N} + e^+ + \nu_e$ & \\
     & $^{17}$F   &$^{17}{\rm F} \to ^{17}{\rm O}+e^+ +\nu_e$ & \\
     \\
     Others & $^{18}$F neutrinos & $^{18}{\rm F} \to ^{18}{\rm O}+e^+ +\nu_e$ & Relevant during He burning\\  \\
     \colrule
     \multicolumn{4}{c}{Thermal reactions}  \\
     \colrule
     \multicolumn{2}{c}{\multirow{2}{*}{Plasmon decay}}  & \multirow{2}{*}{$\gamma^* \to \nu+ \bar{\nu}$}& \multirow{2}{*}{\shortstack{Dominant during early stages of white-dwarf cooling  and\\main thermal process during H-burning for $M\lesssim3M_\odot$.}}\\ \\
     \multicolumn{2}{c}{\multirow{2}{*}{Photoneutrinos$^*$}} & \multirow{2}{*}{$e^- + \gamma \to e^- + \nu_e +\bar{\nu}_e$} & \multirow{2}{*}{\shortstack{Main thermal process during H-burning for $M\gtrsim3M_\odot$\\ and during He burning.}}  \\ \\
     \multicolumn{2}{c}{Pair annihilation$^*$} & $e^+ + e^- \to \nu + \bar{\nu}$ &  Relevant in the late stages of evolution of high-mass stars.\\
     \multicolumn{2}{c}{Bremsstrahlung$^*$} & ${\rm N} + e^- \to {\rm N}+e^- + \nu + \bar{\nu}$& \multirow{2}*{Subdominant} \\
     \multicolumn{2}{c}{Recombination$^*$} & $e^-_{\rm continuum} \to e^-_{\rm bound} + \nu_e + \bar{\nu}_e$ \\
    
     \end{tabular}
    \end{ruledtabular}
    \label{tab:reactions}
\end{table*}

\subsection{Neutrino energy spectrum}
For each time snapshot of  stellar evolution, \texttt{MESA} provides the reaction rates for each thermonuclear process, see Table~\ref{tab:reactions}, as well as the total energy released. The neutrino energy spectrum is not provided as  \texttt{MESA} output, but can be computed  considering the kinematics of each reaction. 

Neutrinos from $\beta$ processes occurring as part of  the pp and CNO cycles produce a continuous spectrum that can be  computed knowing the Q-value of the reaction; whereas reactions with two bodies in the final state produce spectral lines. As for  the thermonuclear component of the neutrino flux, we compute the energy spectrum for all pp-chain reactions (i.e.~the continuous neutrino spectra from pp and hep neutrinos, $^8$B decay, and the spectral lines of pep neutrinos and $^7$Be neutrinos), the CNO-cycle (i.e.~the  energy spectra from the decay of $^{13}$N, $^{15}$O, and $^{17}$F), and the spectrum from $^{18}$F decay. For a  detailed discussion on the modelling of the spectral shape of pp-chain and CNO-cycle reactions, we refer the reader to Ref.~\cite{Vitagliano:2019yzm} and references therein. An analogous procedure approximately holds for $\beta$ processes, such as $^{18}$F.

As for thermal processes, \texttt{MESA} outputs the reaction rates as  functions of radius and for several time steps of the evolution, for plasmon decay, pair annihilation, bremsstrahlung,  photoneutrinos, and neutrinos from recombination. For such processes, the neutrino emission strongly scales as a function of e.g.~the electron temperature and electron number density (see e.g.~Ref.~\cite{Itoh:1996}). Hence, we extract the radial profiles of the relevant thermodynamic quantities from  \texttt{MESA}  at the relevant time snapshots during stellar evolution; relying on such profiles, one can compute the energy spectrum of neutrinos from plasmon decay~\cite{Braaten:1993jw,Haft:1993jt}, pair annihilations~\cite{Dicus:1972yr,Misiaszek:2005ax}, bremsstrahlung ~\cite{Guo:2016vls}, recombination~\cite{Pinaev:1964zem,Kohyama:1993}, and photoneutrinos~\cite{Dutta:2003ny}. As discussed in Sec.~\ref{sec:flux},  we only account for  plasmon decay as contributing to white dwarfs cooling, and we refer the reader to Sec.~\ref{sec:flux} for  details on the relevance of the remaining thermal processes.

In thermonuclear reactions, only electron neutrinos are produced. However, thermal processes produce neutrino-antineutrino pairs. Photoneutrinos and recombination neutrinos are always of electron flavor, while all the other thermal processes produce a mixture of electron-type and non-electron type neutrino pairs, where the ratio is fixed by the structure of  weak interactions. When traveling towards the outer layers of stars, and depending on their energy and the  radial profile of the electron number density at each stellar evolution stage, neutrinos undergo flavor conversion~\cite{Wolfenstein:1977ue,Mikheyev:1985zog}. Flavor conversion in the source, together with  loss of coherence en route to Earth, determines the flavor composition of neutrinos reaching Earth. We do not account for flavor conversion  due to the complexity of integrating the energy-dependent and time-dependent flavor evolution  for each progenitor. Therefore, hereafter we  report the total neutrino and antineutrino flux for all six flavors.

\section{Galactic stellar neutrino flux}
\label{sec:emission}

The GS$\nu$F  at Earth and at present time is: 
\begin{align}
    \Phi_\nu =  \int \text{d}^3 \vec{r}\,&\frac{N_* (\vec{r})}{ 4\pi D_\oplus^2(\vec{r})}\int \text{d}t\,\zeta(t, \vec{r})\,\nonumber \\&\times  \int \text{d}M\,\xi(M) \frac{{\rm d} \mathcal{\phi}_\nu}{{\rm d}E_\nu} (M, \tau)\, . 
    \label{eq:nu-flux}
\end{align}
Here, $N_*$ denotes the number of stars at the location $\vec{r}$ expressed in Galactic coordinates and corresponding to the distance from Earth $D^2_\oplus(\vec{r})$; $\zeta$ denotes the SFH normalized to unity (introduced in Eq.~\ref{eq:sfh}), $\xi$ is the IMF normalized to unity (defined as in Eq.~\ref{eq:imf}) under the assumption of uniform IMF across the Milky Way, $\tau = t_U -t$ with  $t_U = 13.5$~Gyr being the age of the Universe. The neutrino energy flux emitted from a star of ZAMS mass $M$ and age $\tau$ is ${\rm d}\mathcal{\phi}_\nu/{\rm d}E_\nu$. 

For simplicity, we treat the bulge, thin disk, and thick disk separately and we only consider one generation of stars. This allows us to break Eq.~\ref{eq:nu-flux} into three  integrals with independent SFH (cf.~Sec.~\ref{sec:sfh}).
As for the spatial distribution of stars, we construct a finely binned spatial grid for each of the Milky Way components. Under the assumption of isotropic IMF, the number density of stars at each discrete volume element of the 3D grid ($V_k$) is the ratio between the baryon density of each Galactic component (see Table~\ref{tab:MWdensity}) and the mean mass, $\overline{M}$, according to the IMF, i.e.
\begin{align}
    \overline{M} = \frac{\int_{0.08M_\odot}^{100M_\odot}\,{\rm d}M \, \xi(M) M }{\int_{0.08M_\odot}^{100M_\odot}\,{\rm d}M \,\xi(M)}\,.
\end{align}
For each 3D volume element, we draw a random position ($\vec{r}_k$) and assume that all stars in that volume element are located in $\vec{r}_k$. Then, the spatial integral in each of the elements of the spatial grid is
\begin{align}
    \int_{V_k} {\rm d}^3\vec{r} \frac{N_* (\vec{r})}{4\pi D^2_\oplus (\vec{r})} \simeq \frac{1}{4\pi D^2_{\oplus}(\vec{r}_k)} \frac{\int_{V_k} {\rm d}^3 \vec{r}  \rho_{n}(\vec{r})}{\overline{M}}  \equiv \varepsilon_{k}^{(n)}\, ,
\end{align}
where $\rho_n$ is the baryon  density of the bulge (Eq.~\ref{eq:bulge}) and each of the disks (Eq.~\ref{eq:disk}). The index $n$ indicates  each of the three components, i.e.~$n = \lbrace$bulge, thin disk, thick disk$\rbrace$.

We also consider a grid of stellar ages for which we compute the average neutrino emission. We use the normalized SFH distribution and IMF as weights:
\begin{align}
    w_{ij} = \frac{\int_{M_{i,{\rm min}}}^{M_{i,{\rm max}}}\, {\rm d}M \, \xi(M) }{\int_{0.08M_\odot}^{100M_\odot}\, {\rm d}M\,\xi(M) } \frac{\int_{t_{j,{\rm min}}}^{t_{j,{\rm max}}}\,{\rm d}t\, \zeta(t_U-t) }{\int_{0}^{t_U}\,{\rm d}t\, {\zeta}(t_U- t) }\, .
\end{align}
Here $M_{i,{\rm min}}$ and $M_{i,{\rm max}}$ indicate the minimum and maximum mass of the $i$-th mass bin. Note that since the bulge,  thin and thick disks have different SFHs (see Sec.~\ref{sec:sfh}), we compute such weights for each of the three Galactic components separately.

Using the discretization method outlined above,  Eq.~\ref{eq:nu-flux} becomes
\begin{align}
\label{eq:SnuB}
    \Phi^{(n)}_{\nu} = \sum_k \varepsilon^{(n)}_{k} \sum_{i,j} w^{(n)}_{ij} \frac{d\mathcal{\phi}_\nu}{dE_\nu}(M_i, \tau_j)\, .
\end{align}
The cumulative neutrino flux is  the sum of the neutrino emission from the three Galaxy components.

In our numerical modeling, we consider $4331$ volume elements for the bulge, and $50601$ volume elements for each of the disks. For low- and intermediate-mass stars, our stellar-age grid has $2700$ elements linearly spaced between $5$ and $13.5$~Gyr. For high-mass stars, we adopt $4000$ bins in stellar age between $5000$~yr and $20$~Myr.
We have performed resolution tests for the stellar age grid and the spatial grid to ensure convergence of our results (plots not shown here).

\section{Results: stellar neutrino flux}
\label{sec:flux}

In this section, we present our findings on the GS$\nu$F  from low-, intermediate- and high-mass stars in our Galaxy. We discuss the reactions dominating the neutrino emission for each class and discriminate among neutrino and antineutrino emission from the bulge,  the thin disk, and thick disk.

\begin{figure*}
    \centering
    \includegraphics[width=\linewidth]{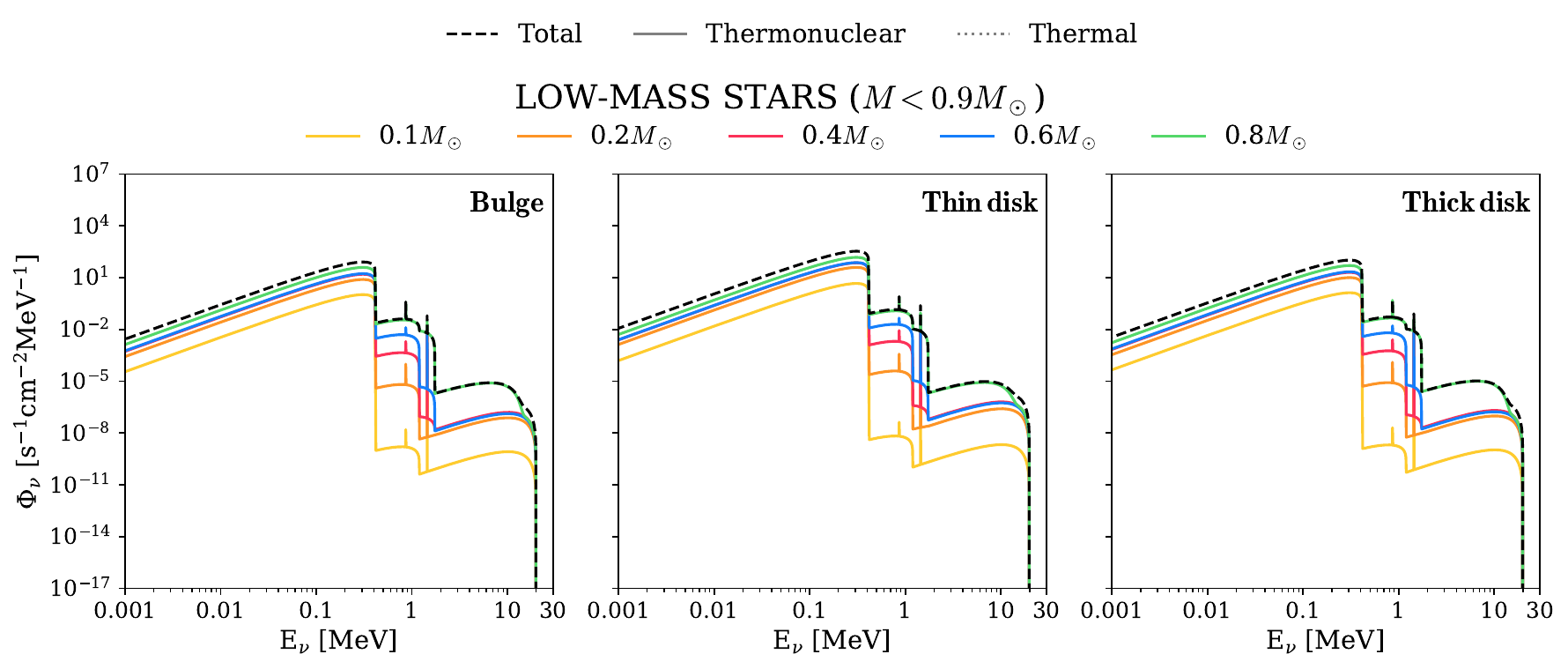}
      \vspace{0.1cm}
    \includegraphics[width=\linewidth]{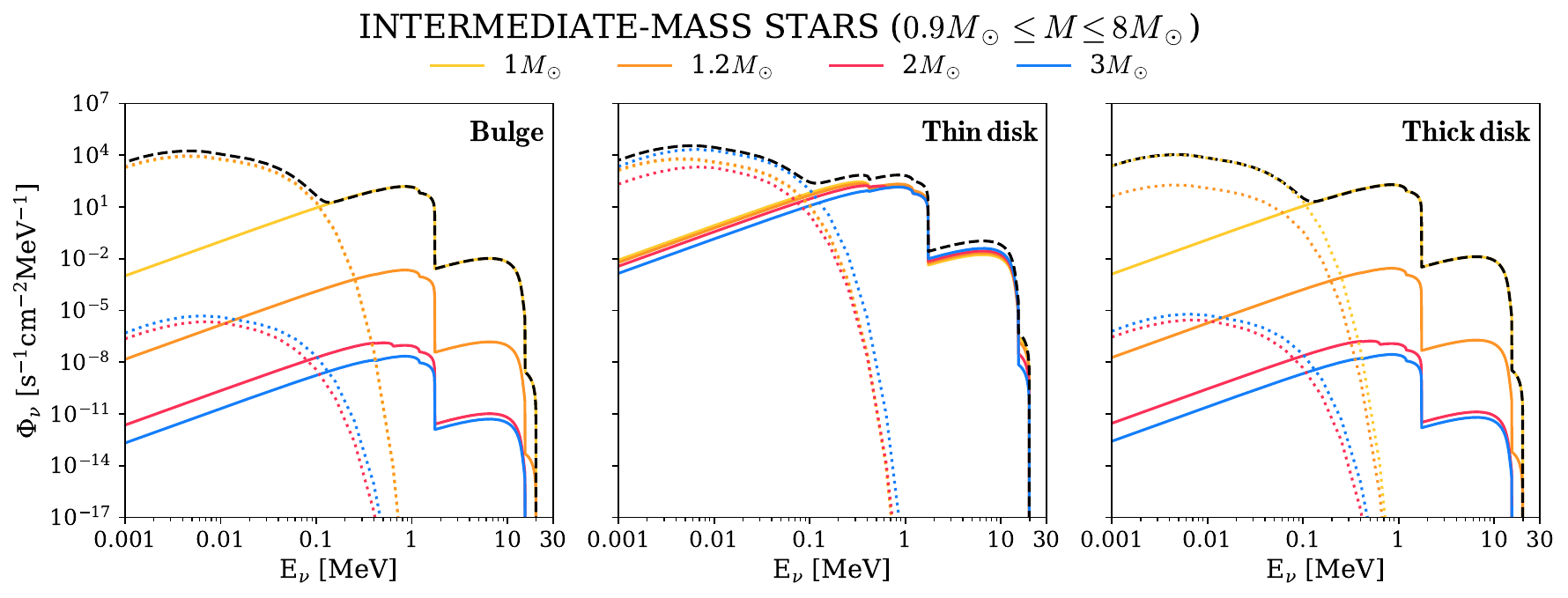}
    \vspace{0.1cm}
\includegraphics[width=\linewidth]{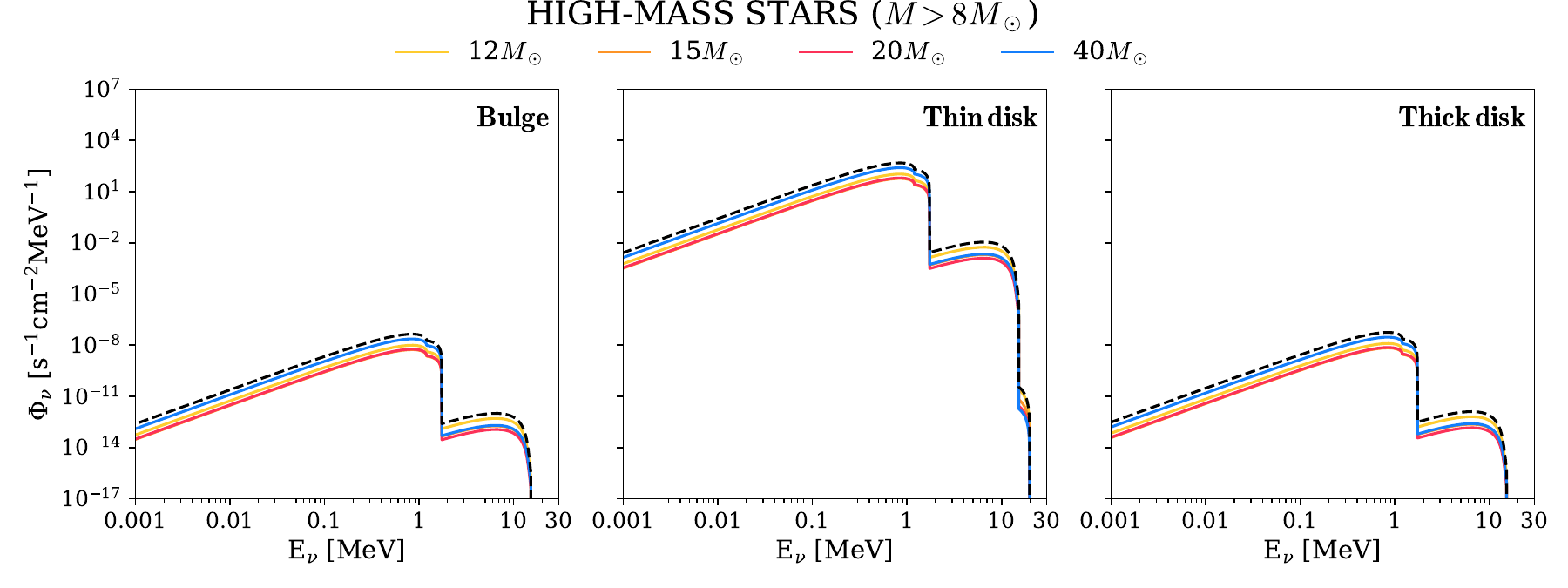}
    \caption{Galactic stellar   neutrino and antineutrino flux at Earth for all flavors and from all low-mass (top), intermediate-mass (middle), and high-mass  (bottom) stars in our Galaxy. The (anti)neutrino emission from the bulge, thin disk, and thick disk is displayed from left to right, respectively. The total  emission for each mass range is marked by a black dashed line; the solid (dotted) lines correspond to thermonuclear (thermal) neutrinos. The (anti)neutrino emission from low- and high-mass stars is dominated by thermonuclear processes, while the low-energy tail of intermediate-mass stars is due to thermal processes. Overall, the largest (anti)neutrino emission comes  from the thin disk, independent of the ZAMS mass range.}
    \label{fig:ranges}
\end{figure*}

\subsection{Low-mass stars}
For low-mass stars (we refer the reader to e.g.~Refs.~\cite{DellOmodarme:2012awu,Bressan:2012zf} for details on the evolution of low-mass stars including post main-sequence stages), we  follow the evolution of stars with ZAMS mass of  $0.1M_\odot$, $0.2M_\odot$, $0.4M_\odot$, $0.6M_\odot$ and $0.8M_\odot$, as outlined in Sec.~\ref{subsec:solar-calibration}. We evolve such models until the terminal-age main sequence, i.e.~until hydrogen is close to being exhausted and the central mass fraction of $^1$H reaches $0.01\%$. We halt the evolution of these models before the end of the main sequence because the lifetime of these stars as main sequence stars would be larger than the age of the Universe. We then compute the neutrino emission by means of Eq.~\ref{eq:SnuB}. Our results concerning the evolution of low-mass progenitors is in agreement with the existing literature~\cite{Farag:2023xid}.

The top panel of Fig.~\ref{fig:ranges}  displays the GS$\nu$F  at Earth from  low-mass stars  in our Galaxy. The neutrino luminosity of  low-mass stars  increases as a function of the ZAMS mass and as the star ages  in its main sequence phase. Therefore, the main contribution to the GS$\nu$F from Galactic low-mass stars comes from the oldest and most massive  stars of this subpopulation. The leading neutrino emission channel is the pp chain, although neutrinos from the CNO cycle start to become relevant towards the end of their main sequence lifetime. Lines emerge from  the continuum spectrum in Fig.~\ref{fig:ranges}; such spectral lines  are characteristic of the $^7$Be and pep reactions.

The GS$\nu$F  from  low-mass stars shown in the top panel of Fig.~\ref{fig:ranges}  is divided according to the contribution from the bulge, thin and thick disks, from left to right, respectively.
The emission from the thin disk dominates since the latter has the largest stellar population. Notice, however, that a bump is expected in the spectrum at $1$--$1.6$~MeV from the CNO cycle and due to the contributions from the bulge and thick disk, as visible  from Fig.~\ref{fig:spectrum} (orange curve). The bulge and the thick disk  are the oldest in the Milky Way, therefore a fraction of their stellar population has already reached the temperature threshold for the  CNO cycle to become a relevant neutrino production channel. 

\subsection{Intermediate-mass stars}
\label{sec:intermediate}
\subsubsection*{Main neutrino production channels}
Our sample of intermediate-mass stars consists of progenitors with ZAMS mass $M = 1 M_\odot$, $1.2 M_\odot$, $2 M_\odot$, $3 M_\odot$. We explore the evolution of these stellar models employing  {\tt MESA} and then compute the neutrino energy spectrum for this population, as outlined in Sec.~\ref{subsec:solar-calibration} and Eq.~\ref{eq:SnuB}. For these stars, the CNO cycle dominates the hydrogen burning for $M> 1.2M_\odot$, and is of relevance for lower masses. As for low-mass stars, the neutrino luminosity  increases  as a function of the ZAMS mass and stellar age during the main sequence.   

During the post-main-sequence evolution, since the core temperature of these stars is not large enough ($\lesssim 10^9$~K) to fuse carbon, an inert core of carbon and oxygen builds up. The star then loses its outer layers to stellar winds, forming a planetary nebula and a white dwarf. White dwarfs lack a source of heat; electron degeneracy supports them against gravitational collapse. Hence, first they cool down by means of neutrino emission (mainly via plasmon decay, which we compute following Ref.~\cite{Braaten:1993jw}); then, at later stages,  radiative cooling takes over. We have tested that our stellar models are in  agreement with  existing literature, including recent  work focusing on neutrino emission~\cite{Farag:2020nll}.

The middle panels of Fig.~\ref{fig:ranges} show the GS$\nu$F  from intermediate-mass stars in our Galaxy.  
For the oldest Galactic components   (the bulge and the thick disk), the neutrino emission from the smaller ZAMS masses is the largest, since these are the only stars that are still in the main sequence to date. The thermal neutrino emission from the cooling of white dwarfs is also led by intermediate-mass stars with smaller mass due to their abundance, as implied by the IMF and shown in Fig.~\ref{fig:model}. For the thin disk, the contribution from thermonuclear reactions is similar for all masses in this range, since there is a significant part of the population in the main-sequence stage. These differences arise from the relative importance between the pp chain and the CNO cycle during hydrogen burning according to the  core temperature. The thermal part of the neutrino energy spectrum is coming from  plasmon decay and it is dominated by the highest ZAMS masses in the low-energy range.

\subsubsection*{Additional neutrino production channels}

At the end of the main sequence, stars with mass  $\lesssim 2\ M_\odot$ have a degenerate core. When the mass of the degenerate core reaches $\sim 0.5\ M_\odot$, the core temperature is about $10^8$~K and helium ignition is activated via triple-$\alpha$ process. Helium ignition is responsible for  an electromagnetic flash.
The helium flash is sometimes followed by subflashes; such flashes  become  less likely as the progenitor mass increases and  burns a smaller amount of helium. As a consequence,  the nuclear energy that is generated mainly goes into lifting  electron degeneracy in the core. This process leads to the formation of a stable
convective core of helium burning under non-degenerate conditions.

The slowest step in the CNO cycle is proton capture on $^{14}$N. As a result, when hydrogen burning is completed, one expects a significant abundance of $^{14}$N. During  helium ignition, $^{14}$N is converted into $^{22}$Ne by the following set of reactions,
\begin{subequations}
    \begin{align}
    &^{14}{\rm N} \,+\, \alpha \to \gamma \,+\,  ^{18}{\rm F}\, ,\\
    &^{18}{\rm F} \to e^+ + \mathbf{\nu_e} \,+ \,^{18}{\rm O}\, ,\\
    & ^{18}{\rm O} \,+\, \alpha \to \gamma \,+\,^{22}{\rm Ne}\, .
\end{align}
\end{subequations}
Hence, the helium flash (and the subflashes) are correlated with the emission of a burst (or multiple bursts) of neutrinos from $^{18}$F decay (with a Q-value of $1.655$~MeV)~\cite{Serenelli:2005nh}. Despite their large neutrino luminosity, these flashes barely affect the time-integrated  energy emitted in neutrinos from each intermediate-mass star during its life. In addition, the end point of the energy spectrum of neutrinos from $^{18}$F is very close to the one of neutrinos from $^{15}$O decay ($E_\nu \simeq 1.73$~MeV) from the CNO cycle. Hence, the neutrino energy spectra from these two channels tend to overlap in energy.

For masses $M > 2\ M_\odot$, helium ignition occurs in the core  under non-degerate conditions. Hence, no helium flash (nor neutrino burst) is expected.

At later stages in the post-main-sequence evolution,  during the asymptotic giant branch phase, hydrogen and helium burning occur in geometrically thin shells
on top of the carbon-oxygen core. Thermal pulses occur when the helium shell  ignites. These episodes also produce
a burst of neutrinos from $^{18}$F decay. Such neutrino bursts  are less luminous than the ones following helium flashes. We stress, however, that this phase of stellar evolution is very uncertain, therefore existing models are  calibrated to observational data~\cite{Pastorelli:2019a,Pastorelli:2020a}.
Importantly, the number of thermal pulses depends on the treatment of mass loss~\cite{Herwig:2005zz}. The latter may halt pulses once the envelope has been ejected.

In our models, a parametric treatment of convection has been adopted, based on mixing-length theory, and calibrated to match observations in the Sun. However, for masses larger than $1.5$--$2\ M_\odot$, stars develop a convective core and a radiative envelope, contrary to the case of solar-mass stars. Hence, the number of thermal pulses in our models, and even their existence, might be a result of the approximations intrinsic to our model. 
Despite that, given the short duration of such pulses, their contribution to the GS$\nu$F is negligible.

During the hydrogen- and helium-burning stages of evolution, the stellar-core temperature is  such that neutrinos from plasmon decay and photoneutrinos are also emitted. The neutrino flux from these reaction channels was estimated to be comparable or smaller than the number of neutrinos emitted during white-dwarf cooling~\cite{Brocato:1997tu}. Our forecast of the GS$\nu$F  does not account for these thermal processes in these stages of stellar evolution; in this sense, our prediction is conservative. These thermal processes would partially  modify the thermal component of the spectrum (below $100$~keV)  in shape, enhancing it. Given the strong dependence of these processes on temperature, a detailed assessment of their impact is beyond the scope of this work; in fact,  a more accurate treatment of convection in the core would be needed to this purpose. 

\subsection{High-mass stars}
\subsubsection*{Main neutrino production channels}
Our discrete sample of high-mass stars, evolved using {\tt MESA}  as outlined in Sec.~\ref{sec:evolution}, consists of four progenitors with ZAMS mass  of $12\ M_\odot$, $15\ M_\odot$, $20\ M_\odot$, and $40\ M_\odot$. We follow the evolution of these progenitors until they reach conditions close to core collapse--namely when the material inside the iron core exceeds $10^7$~cm/s.

During their life as main sequence stars, neutrino emission is dominated by the CNO cycle. At later stages, thermal neutrino emission, mainly from pair processes becomes relevant.  
This efficient  energy-loss mechanism sets a rapid
evolution  (years to hours)  of the pre-supernova stars during the advanced stages
of nuclear fusion. Hence,  neutrino emission during the main-sequence evolution dominates the contribution to the GS$\nu$F.

The bottom panels of Fig.~\ref{fig:ranges} show the GS$\nu$F from high-mass stars, computed based on Eq.~\ref{eq:SnuB}.
Despite the smaller abundance of high-mass stars dictated by the  IMF (cf.~Fig.~\ref{fig:model}), the  neutrino emission grows with the ZAMS mass in the range below $1.6$~MeV and is mainly due to the CNO cycle.
For higher neutrino energies, the spectrum is shaped by pp-neutrinos, $^8$B  and hep neutrinos, from progenitors with mass $\lesssim 15 M_\odot$ within this population, i.e.~from the less massive stars of this subpopulation.

High mass stars have a very short main-sequence lifetime. Therefore, the contribution to the GS$\nu$F is smaller for the older components of the Milky Way (i.e., the bulge and thick disk). However, since star formation in the thin disk is a continuous process, occurring even in recent times, we find that  there is a significant population of high-mass stars in the main-sequence stages of evolution at present; this population of main-sequence stars  emits a copious amount of neutrinos.

We note that our  treatment of convection calibrated to the Sun might not be accurate for high-mass stars with a convective core. This could affect  the late stages of stellar evolution. However, such stages have a negligible impact on the  GS$\nu$F due to their short duration and overall  neutrino emission smaller than the one integrated during the remaining lifetime of the massive star.

\subsubsection*{Additional neutrino production channels}

As the pre-supernova phase approaches,  $\beta$-processes contribute to increase the electron-to-baryon ratio, making the medium more neutron rich and favoring the emission of a significant amount of MeV neutrinos~\cite{Odrzywolek:2010zz,Yoshida:2016imf,Kato:2017ehj,Patton:2017neq}. 
Thermal neutrino emission from the pre-supernova phase should be expected~\cite{Misiaszek:2005ax,Kato:2020hlc}. 
Nevertheless, we do not account for the contribution of pre-supernova neutrinos to the GS$\nu$F. This is justified because about $30$ supernova candidates have been identified at a distance smaller than $1$~kpc from Earth~\cite{Mukhopadhyay:2020ubs}; their pre-supernova emission is expected to increase substantially only less than a day before the core collapse of the massive star. As a consequence, we estimate that the cumulative neutrino emission from  these local supernova candidates is negligible with respect to the total neutrino emission from the remaining massive stars in our Galaxy.

Notice that  our approach yields a local rate of $2$--$3$ stars with $M \gtrsim 8 M_\odot$ undergoing core collapse per century; this is  compatible with more sophisticated estimates~\cite{Rozwadowska:2020nab}. Since the number of core-collapse supernovae  in the Galaxy at present is $\ll 1/$s, we conclude that supernovae do  not contribute to the GS$\nu$F. Instead, a local core-collapse supernova would be observed at neutrino telescopes as a transient signal~\cite{Mirizzi:2015eza,Tamborra:2024fcd,Raffelt:2025wty}.

For high-mass stars, it is known that photoneutrino production is the dominant thermal process during main-sequence evolution, especially during the helium burning phase. Previous work evaluated this contribution to be smaller than or comparable to the one resulting from the cooling of a white dwarf~\cite{Brocato:1997tu}. For  reasons analogous to the ones  discussed in Sec.~\ref{sec:intermediate}, we do not include this contribution in our prediction. 

\subsection{Cumulative neutrino flux}
\begin{figure*}
    \centering
    \includegraphics[width=\linewidth]{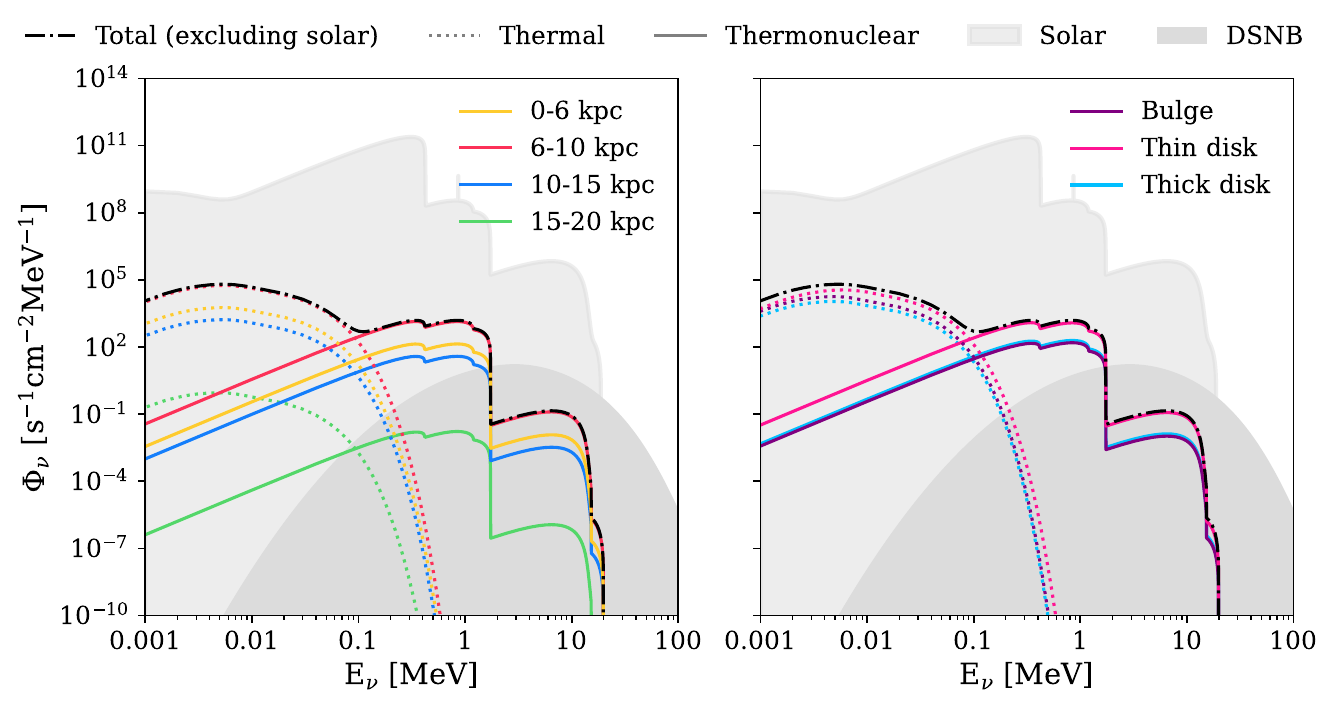}
    \caption{Galactic stellar  flux of neutrinos and antineutrinos (black dashed line) as a function of the neutrino energy and for all flavors. The solid and dotted lines distinguish between  the thermonuclear or thermal origin of  neutrino emission, respectively. The flux from solar neutrinos and the DSNB, modeled as in Ref.~\cite{Vitagliano:2019yzm}, are shown as light and dark  gray bands, respectively. {\it Left:} Contributions to the GS$\nu$F for different distances from Earth. We distinguish among $5$ distance bins in the range between $0$ and $20$~kpc from Earth ($0$--$6$, $6$--$8$, $8$--$10$, $10$--$15$, and $15$--$20$~kpc plotted in orange, red, blue, and green, respectively).
    The GS$\nu$F  is dominated from stars  between $6$ and $10$~kpc from Earth.
    {\it Right:}  Contributions to the GS$\nu$F  from the bulge, the thin disk, and the thick disk are plotted in purple, magenta, and light blue, respectively. Stars in the thin disk of the Milky Way provide the largest contribution to the GS$\nu$F.}
    \label{fig:dist_comp}
\end{figure*}

The GS$\nu$F  is displayed in Fig.~\ref{fig:spectrum}. For readers wishing to reproduce their own version of the GS$\nu$F plot,  we provide the tabulated fluxes of the different GS$\nu$F components displayed in Fig.~\ref{fig:spectrum} as  Supplemental Material of this paper (and ancillary data files in the arXiv repository). In addition, Fig.~\ref{fig:dist_comp}  highlights the contributions to the GS$\nu$F  according to the  distance of stars  from Earth and the overall contributions from the bulge, thin and thick disks (in the left and right panels, respectively). For energies smaller than $100$~keV, the GS$\nu$F is dominated by thermal processes, in particular plasmon decay from  white dwarfs marking the end of the evolution of  intermediate-mass stars; see also Fig.~\ref{fig:ranges}.
For larger energies, thermonuclear processes dominate the GS$\nu$F. Overall, the flux peaks at energies  $\lesssim 0.5$~MeV; this emission is mostly due to  pp neutrinos produced in low- and intermediate-mass stars. For energies up to $\sim 1.6$~MeV, the GS$\nu$F is shaped by  CNO reactions occurring in the core of intermediate- and high-mass stars. 
Therefore, the spectral lines from $^7$Be and pep neutrinos displayed  in Fig.~\ref{fig:spectrum} for low-mass stars (orange curve)  do not leave any trace in the  overall emission plotted as black dashed lines in Fig.~\ref{fig:dist_comp}. The GS$\nu$F  is  significantly reduced at high energies ($\gtrsim 1.5$~MeV), with a spectral shape determined by  $^8$B and hep neutrinos.

The left panel of Fig.~\ref{fig:dist_comp} suggests that  stars within  $6$ and  $10$~kpc from  Earth are the main contributors to the GS$\nu$F. Notably, this distance range  includes the Galactic Center (located at roughly $8$~kpc from Earth), where the number density of stars is the largest.

The right panel of Fig.~\ref{fig:dist_comp} highlights that the largest contribution to the GS$\nu$F  comes from the thin disk of the Milky Way. The reason is twofold. First, the thin disk is  the most massive component (and hence the one with the largest number of stars,  under the assumption of isotropic IMF). Second, the thin disk  is the youngest component of the Galaxy and therefore the one hosting the largest number of intermediate and high-mass stars in the main sequence  (with the main sequence being the  evolutionary stage dominating the neutrino emission).

Figures~\ref{fig:spectrum} and \ref{fig:dist_comp} suggest that the neutrino flux from stars in our Galaxy is  about $5$ orders of magnitude smaller than the solar neutrino flux and the diffuse supernova neutrino background (DSNB). 
Although the development of  dedicated experimental techniques would be necessary to make the detection of stellar neutrinos possible, it is worth noting that the magnitude of the GS$\nu$F  has never been larger than at present. The Earth  formed when the Universe was about $4.5$~Gyr old (after the first infall leading to star formation in the bulge and thick disk);   star formation has taken place at a slightly decreasing rate since then. Hence, one can expect that the population of intermediate- and high-mass stars in the thin disk (which  happens to be the most massive component of the Galaxy) has  increased since then. Most of these are main-sequence stars, with neutrinos being abundantly produced.

Our forecast of the GS$\nu$F is in  overall agreement with earlier work presented in Refs.~\cite{Brocato:1997tu,Porciani:2003zq}, both as for the spectral shape  and normalization of the flux. However, our work is based on up-to-date 1D modelling of stellar evolution (with a complete nuclear reaction network) and accounts for a modeling of the Milky Way stellar population based on recent observational data. 
In principle, one could extend these results and estimate the diffuse cumulative flux from  stars in other galaxies. Estimates of this sort can be found in Refs.~\cite{Brocato:1997tu,Porciani:2003zq} based on simple assumptions. 

\subsection{Components of the GS$\nu$F}
\begin{table}[]
    \centering
    \caption{Solar neutrino flux~\cite{Gonzalez-Garcia:2023kva} and GS$\nu$F (excluding the solar flux) from the main thermonuclear reactions. For each reaction channel, the $Q$ value (maximum $\nu_e$ energy for the continuum spectra)  is also provided.}
    \renewcommand{\arraystretch}{1.4}
    
    \begin{ruledtabular}
    \begin{tabular}{lccc}
    \multirow{2}*{\begin{tabular}[c]{@{}c@{}}Process\\{}\end{tabular}} & \multirow{2}{*}{\begin{tabular}[c]{@{}c@{}}Sun\\ {[}cm$^{-2}$s$^{-1}${]}\end{tabular}} & \multirow{2}{*}{\begin{tabular}[c]{@{}c@{}}GS$\nu$F (w/o Sun)\\ {[}cm$^{-2}$ s$^{-1}${]}\end{tabular}} & \multirow{2}{*}{\begin{tabular}[c]{@{}c@{}}$Q$ [MeV]\\{}\end{tabular}}\\
    & & & \\
    \colrule
        pp & 5.94$\times10^{10}$  &$2.38\times 10^2$   & 0.420  \\
        \multirow{2}*{$^{7}$Be}& \multirow{2}*{$4.93\times 10 ^9$} & \multirow{2}*{25.8} & 0.862 (89.7\%) \\
         & & & 0.384 (10.3\%) \\
        pep & $1.42\times 10^{8}$& 0.60& 1.442\\
        $^8$B & $5.20\times 10^6$& 1.02 & 15.2\\
        hep & $3.0\times10^4$& $4.6\times 10^{-5}$& 19.795\\
        $^{13}$C & $3.5\times 10^8$ & $5.37 \times 10^2$ & 1.199	\\
        $^{15}$N & $2.5\times 10^8$ & $5.40\times 10^2$  & 1.732 \\
        $^{17}$F & $5.5\times 10^7$ & 2.10 & 1.740 \\
        $^{18}$F & - & 30.2& 0.644\\ 
    \end{tabular}
    \end{ruledtabular}
    
    \label{tab:fluxes}
\end{table}
In order to compare the various components of the GS$\nu$F to the well-known solar neutrino flux, Table~\ref{tab:fluxes} summarizes the observed flux of neutrinos in all flavors from the thermonuclear reactions taking place in the  Sun and the GS$\nu$F. We note that instead of relying on the solar neutrino flux from Ref.~\cite{Vitagliano:2019yzm} (as otherwise done in this work), we  quote the solar neutrino components from Ref.~\cite{Gonzalez-Garcia:2023kva}, obtained  fitting  all available solar neutrino data. For completeness, we also  provide the $Q$ value for each thermonuclear reaction channel, which corresponds to the maximum electron neutrino energy for the reaction channels leading to  spectral energy distributions  and  the neutrino energy for the $^7$Be and pep  channels responsible for spectral lines.
We can see that the solar neutrino flux dominates over the GS$\nu$F  across all neutrino-production channels.
However, the ratio of the flux from pp neutrinos to CNO-cycle neutrinos is larger for the Sun than for the rest of the GS$\nu$F. This is because the neutrino emission from intermediate- and high-mass stars is dominated by the burning of hydrogen through the CNO cycle during their life as main sequence stars. 

\begin{figure*}
    \centering
    \includegraphics[width=\linewidth]{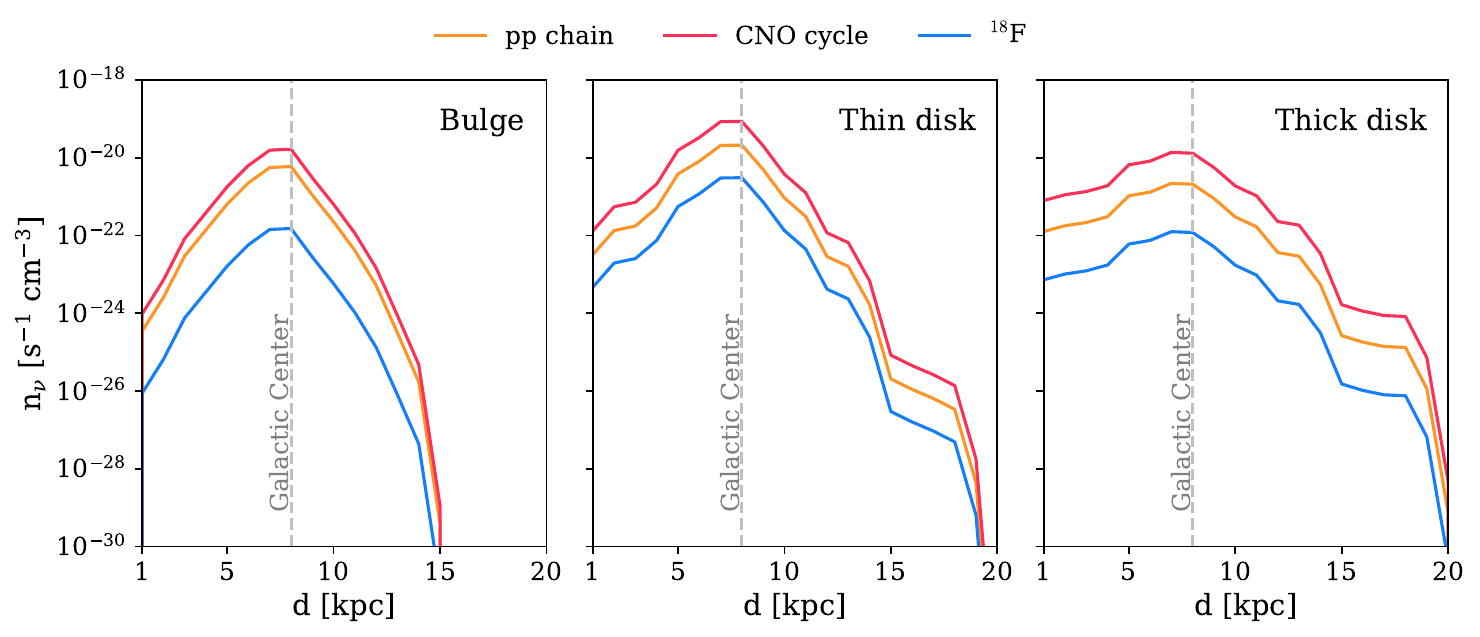}
    \caption{Differential neutrino emission rate from the thermonuclear reactions contributing to the GS$\nu$F  as a function of the distance from Earth from the bulge, thin disk, and thick disk in the left, middle, and right panels, respectively. The orange, red, and blue lines correspond to the flux from the pp chain, the CNO cycle, and $^{18}$F neutrinos from He burning flashes (see main text for details). The distance of the Galactic Center from Earth is marked by a gray dashed vertical line.  The flux from the CNO cycle dominates, with the largest contribution coming from the thin disk.}
    \label{fig:reactions}
\end{figure*}

Figure~\ref{fig:reactions} shows the rate per unit volume  of  pp, CNO, and $^{18}$F neutrinos as functions of the distance from Earth for the bulge, the thin disk, and the thick disk. For all Galactic components  CNO neutrinos constitute the largest contribution to the overall neutrino flux. However,  the radial evolution of each contribution depends  on the matter distribution of each of the spatial components of our Milky-Way model. The largest neutrino flux at Earth should be expected from the Galactic Center, as highlighted by the  vertical dashed line at $8$~kpc. Of course, if the observer were to point in other directions, the GS$\nu$F should be expected to be  significantly smaller. 

\begin{figure*}
    \centering
    \includegraphics[width=\linewidth]{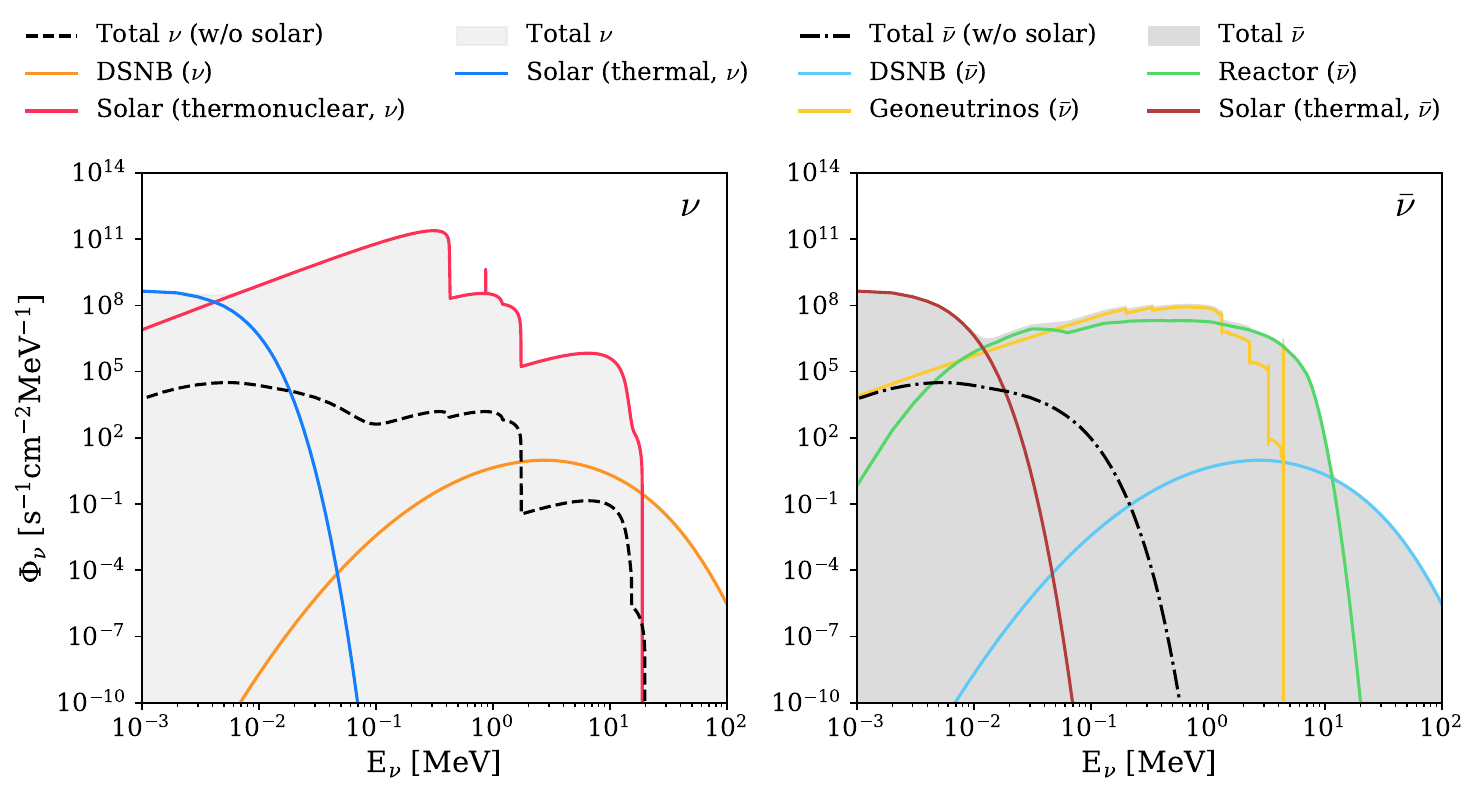}   \caption{Galactic stellar flux of neutrinos (left panel, dashed line) and antineutrinos (right panel, dash-dotted line) as  functions of the neutrino energy. The shaded region in the left (right) panel indicates the total neutrino (antineutrino) flux, including background sources to the GS$\nu$F (i.e., solar neutrinos, DSNB, geoneutrinos, and reactor antineutrinos from~Ref.~\cite{Vitagliano:2019yzm}). In the left panel, the  DSNB neutrinos as well as the solar themonuclear and thermal neutrinos are show in  orange, red, and dark blue, respectively. In the right panel, the DNSB antineutrinos, reactor antineutrinos, geoneutrinos, and themal solar antineutrinos are displayed in light blue, green, yellow, and brown, respectively. 
    Notice, that the fluxes of reactor antineutrinos and geoneutrinos are strongly location dependent.
    \label{fig:nu-antinu}}
\end{figure*}

Figure~\ref{fig:nu-antinu} represents the GS$\nu$F contributions of neutrinos and antineutrinos. The high-energy tail of the GS$\nu$F, for  $E_\nu \gtrsim 0.1$~MeV, comes from thermonuclear reactions and, therefore, is composed solely of neutrinos. However, the thermal component of the flux has equal contributions from neutrinos and antineutrinos since these are produced in pairs in thermal processes. For comparison, we also show other relevant neutrino and antineutrino backgrounds in the energy range of interest, such as the DSNB, solar neutrinos, geoneutrinos, and reactor antineutrinos~\cite{Vitagliano:2019yzm}. We stress, however, that the reactor antineutrinos and geoneutrinos, although could be backgrounds to the GS$\nu$F detection, are strongly dependent on the neutrino telescope location; hence the fluxes show in Fig.~\ref{fig:nu-antinu} should only serve for orientation.

\section{Concluding remarks}
\label{sec:conclusion}

Stars produce neutrinos throughout their life because of thermal and thermonuclear processes. In addition to the routinely observed solar neutrinos, we should expect the GS$\nu$F  between a few keV and tens of MeV from all stars in our Galaxy. In this work, we compute this neutrino signal. In order to do so, we consider three different components of our Galaxy: the bulge, a thin disk, and a thick disk. The SFH for each of these components  is modelled according to the two-infall model.
The  baryon density distribution is inferred from the analysis presented in Ref.~\cite{Cautun:2020} which uses Gaia Data Release 2. 
In order to compute the neutrino energy spectrum expected from each star, we construct a suite of $13$ progenitors with solar metallicity and ZAMS  mass ranging between $0.1 M_\odot$ and $40M_\odot$. We rely on  the stellar evolution code \texttt{MESA} to evolve each of these stars from their pre-main sequence phase to their final fates. We assume that such models are representative of the Galactic population of stars with ZAMS  in the range between $0.08M_\odot$ and $100M_\odot$. For each time snapshot during  stellar evolution, \texttt{MESA} outputs the reaction rates and total energy released from the thermonuclear (pp chain, the CNO cycle, and $^{18}$F decay) and thermal processes (plasmon decay, photoneutrinos, pair annihilation, recombination, and bremsstrahlung) taking place in the core of each star. We then compute the neutrino energy spectrum for each of these processes during the evolution of the models in our suite.

We find that the bulk of the GS$\nu$F  comes from thermal  processes for neutrino energies $\lesssim \mathcal{O}(0.1)$~MeV and thermonuclear ones otherwise. For energies in the range of $0.1M_\odot$--$0.5$~MeV and above $\sim1.6$~MeV, the GS$\nu$F is dominated by pp neutrinos from stars with ZAMS  smaller than $8M\odot$; the remaining (intermediate) part of the spectrum for $0.5 ~{\rm MeV} \lesssim E_\nu \lesssim 1.6$~MeV results from the CNO cycle occurring in stars with mass $\gtrsim 1 M_\odot$. The neutrino emission mainly comes from the thin disk of the Milky Way, since this is the youngest and most massive component of our Galaxy, with the largest density of stars with ZAMS  mass $\gtrsim 1 M_\odot$ in their main-sequence evolutionary stage. The GS$\nu$F  dominated by thermonuclear reactions is  solely composed of neutrinos, whereas the thermal component of the GS$\nu$F has an equal number of neutrinos and antineutrinos (see Table~\ref{tab:reactions}). The bulk of the GS$\nu$F  comes from stars within $5$ and $10$~kpc from Earth; this implies that we are favorably located to  receive a copious amount of neutrinos from the Galactic Center where the density of stars is the highest. 

The GS$\nu$F lies in the same energy region of solar neutrinos, reactor antineutrinos, geoneutrinos, the DSNB, and possibly the diffuse background of neutrinos from Thorne-{\.Z}ytkow Objects~\cite{Vitagliano:2019yzm,Martinez-Mirave:2025pnz}, although the spectral features of these fluxes are very different and the reactor and geoneutrino contributions are strongly dependent on the neutrino telescope location. In particular, the detection of the GS$\nu$F  is challenged by the fact that this flux is about five orders of magnitude smaller than the one of solar neutrinos.  
It would be crucial to devote  experimental efforts to overcome  solar neutrinos and the DSNB, favoring a unique discovery for stellar archaeology. In fact, the detection of the GS$\nu$F would provide unique insight on the Galactic stellar population, especially for  stars with ZAMS  mass greater than $1M_\odot$.

Leveraging on the directionality of the neutrino  events detected at Earth could be crucial to discriminating between the GS$\nu$F and the solar and supernova backgrounds. In fact, we could combine our precise understanding of the solar neutrino flux with  directional information on solar neutrino events to tag and remove the solar background. Moreover, the annual modulations of the solar neutrino flux could be employed as an additional discriminant factor to disentangle the solar neutrino background from the GS$\nu$F. We expect that a large fraction of GS$\nu$F comes from the Galactic Center. 
However, even with perfect directional reconstruction, the DSNB  might only be reduced by approximately one order of magnitude (at most by a factor $1/4 \pi$ due to the DSNB isotropy). Hence, work is strongly encouraged to improve the detection prospects of neutrinos with  energy $\lesssim 1.6$~MeV, where the DSNB  is negligible.

As a final remark, we stress that a detailed knowledge of the GS$\nu$F  can be of relevance to the discovery of  physic scenarios beyond the  Standard Model. For instance, in the presence of non-standard neutrino self-interactions, neutrinos from distant astrophysical sources would scatter off the GS$\nu$F, leading to potential observable features, such as spectral distortions or changes in the flavor composition. We refer the reader to Ref.~\cite{Berryman:2022hds}--and references therein--for analogous examples  involving astrophysical neutrinos scattering  off the cosmic neutrino background.
Moreover, a high-energy neutrino flux could originate from the scattering of cosmic rays on the GS$\nu$F or the diffuse flux from stars in other galaxies, as  proposed for cosmological relic neutrinos~\cite{Hara:1980mz,Hara:1980b}. If neutrinos were unstable, their emission from stars and decay en route to Earth  could give rise to a detectable flux of photons or dark matter; this scenario was explored  in Ref.~\cite{Porciani:2003zq} based on their prediction of the diffuse flux from all galaxies.

\begin{acknowledgments}
We are grateful to Fabio Iocco, Georg Raffelt, Aldo Serenelli, Alejandro Vigna-G\'omez, Edoardo Vitagliano, Achim Weiss, and Michael Wurm for insightful discussions and comments on our manuscript. This project has received support from the Villum Foundation (Project No.~13164) and the  German Research Foundation (DFG) through the Collaborative Research Center ``Neutrinos and Dark Matter in Astro- and Particle Physics (NDM),'' Grant No.~SFB-1258-283604770.   The Tycho supercomputer hosted at the SCIENCE HPC center at the University of Copenhagen was used to support the numerical simulations presented in this work. 
\end{acknowledgments}

\bibliographystyle{apsrev4-2}
\bibliography{bibliography}

\end{document}